\documentclass[aps,prl,twocolumn,floatfix]{revtex4} 
\usepackage{epsfig}

\begin{document}

\title{\textsc{Constraints on Braneworld Gravity Models from a 
Kinematic Limit on the Age of the Black Hole XTE~J1118$+$480}}

\author{Dimitrios Psaltis} \affiliation{ Departments of
Physics and Astronomy, University of Arizona, Tucson, AZ 85721}

\begin{abstract}
In braneworld gravity models with a finite AdS curvature in the extra
dimension, the AdS/CFT correspondence leads to a prediction for the
lifetime of astrophysical black holes that is significantly smaller
than the Hubble time, for asymptotic curvatures that are consistent
with current experiments. Using the recent measurements of the
position, three-dimensional spatial velocity, and mass of the black
hole XTE~J1118$+$480, I calculate a lower limit on its kinematic age
of $\ge 11$~Myr (95\% confidence). This translates into an upper limit
for the asymptotic AdS curvature in the extra dimensions of
$<0.08$~mm, which significantly improves the limit obtained by
table-top experiments of sub-mm gravity.
\end{abstract}


\maketitle

%%%%%%%%%%%%%%%%%%%%%%%%%%%%%%%%%%%%%%%%%%%%%%%%%%%%%%%%%%%%%%%%%%%%%

The existence of large extra dimensions promises to offer a solution to
the hierarchy problem in physics~\cite{led}. If the standard model is
restricted to act only on a (3+1)-dimensional brane, whereas gravity is
allowed to propagate in the higher-dimensional bulk, the effective Planck
scale in the four-dimensional spacetime can be made significantly larger
than the electroweak scale, matching the experimental requirements.

In a braneworld model, even though it is only gravity that feels the
presence of the extra dimensions, there are still a number of
detectable effects on our (3+1) brane that can be used in constraining
the properties of the bulk. The most direct implication of braneworld
gravity models with large extra dimensions is the deviation of the
inverse-square law for gravity at small scales. Torsion-balance
experiments have been used successfully in rejecting the possibility
of deviations of order unity from Newtonian gravity at scales as small
as $\sim 0.2$~mm~\cite{TBE}.  

Significantly tighter bounds on the higher-dimensional Planck scale
can be obtained by comparing model predictions to a number of
astrophysical phenomena~\cite{kanti}. Unfortunately, such bounds
typically depend on the particular interpretation of astrophysical
observations and suffer from large systematic effects. However, if a
compelling case can be made for any of these phenomena, a very tight
bound on braneworld models can be achieved.

Among the astrophysical tests, the one that requires the least amount
of information is related to the evaporation of black holes. Within
general relativity, the evaporation timescale of a black hole of any
astrophysical mass ($\sim 1-10^9 M_\odot$) is much longer than the age
of the universe. However, application of the AdS/CFT duality in
braneworld gravity models that are asymptotically AdS (such as the
Randall-Sundrum models) shows that this timescale can be significantly
reduced, if the higher-dimensional Planck scale is not much larger
than the electroweak energy scale~\cite{emp}. In particular, the
evaporation time is given by~\cite{emp}
\begin{eqnarray}
\tau&\simeq& 1.2\times 10^2\left(\frac{M}{M_\odot}\right)^3
   \left(\frac{L}{1~\mbox{mm}}\right)^{-2}~\mbox{yr}\;,
\label{eq:life}
\end{eqnarray}
where $M$ is the mass of the black hole and $L$ is the asymptotic AdS
radius of curvature of the bulk.

Because of the no-hair theorem, this theoretical prediction depends
only on assumptions regarding the braneworld model and in particular
on the validity and implementation of the AdS/CFT correspondence. The
basis of the calculation discussed above has been recently
challenged~\cite{fitz}, where it was argued that the strongly coupled
nature of the holographic conjecture allows for time-independent,
i.e., non evaporating, black-hole solutions. This issue will, of
course, be resolved when complete black-hole solutions in braneworld
gravity models are derived. However, even in the absence of such
solutions, heuristic arguments based on similarities with black-string
solutions and the thermodynamic properties of black holes indicate
that static solutions will be unstable, at a timescale given by
equation~(\ref{eq:life}).

The lifetime given by equation~(\ref{eq:life}) was calculated for
black holes in vacuum, and therefore, its application to astrophysical
black holes may not be justified. However, it is important to note
that the fast ``evaporation'' of the black holes described
in~\cite{emp} is in fact a classical phenomenon and is simply
equivalent to the radiation of a large number of modes in the bulk,
which is devoid of matter, by construction. Matter external to the
black hole on the brane will affect this result only if its
gravitational field is comparable to that of the black hole. This is
never the case in systems with stellar-mass black holes, for which the
fraction of mass in the accretion flow is
\begin{equation} 
\frac{M_{\rm acc}}{M}\simeq 5\times 10^{-10} \left(\frac{\epsilon}{0.1}\right)
\left(\frac{R_{\rm acc}}{1~\mbox{AU}}\right)^{3/2} 
\left(\frac{M}{10 M_\odot}\right)^{-1/2}\;.
\end{equation}
Here I assumed a free-falling accretion flow of radial extend $R_{\rm
acc}$, accreting at the Eddington critical rate, with a radiation
efficiency $\epsilon$. Clearly, the presence of matter on the brane
external to the black hole will not affect the gravitational
properties of the latter in the bulk.

A final concern arises by the rapid accretion of matter by
astrophysical black holes, which may overwhelm the rate of
evaporation.  For a black hole accreting at the Eddington critical
rate, the rate of evaporation is larger than the rate of accretion, as
long as the mass of the black hole is~\cite{emp}
\begin{equation}
M\le 50.3 \left(\frac{\epsilon}{0.1}\right)^{1/3}
\left(\frac{L}{1~\mbox{mm}}\right)^{2/3}\;.
\end{equation}
For a $10~M_\odot$ black hole, the rate of accretion of matter is
$\simeq 125$ slower than the rate of evaporation and can, therefore,
be neglected.
 
The prediction given by equation~(\ref{eq:life}) does not depend on
unknowns that usually hamper other astrophysical tests. In fact, using
it to place a constraint on braneworld gravity models requires only a
firm lower limit on the age of an astrophysical black hole. In this
{\em Letter\/}, I use the results of recent observations of the black
hole XTE~J1118$+$480 to place a lower limit on its kinematic age and a
lower bound on the asymptotic AdS radius of braneworld gravity
models~\cite{pos}.

XTE~J1118$+$480 is a compact object in a binary system that lies away
from the galactic plane~\cite{bh}. The high ($\ge 6.4~M_\odot$)
measured mass function of the binary secures the identification of the
compact object as a black hole~\cite{bh}. Its position away from the
galactic plane, as well as its large inferred spatial velocity, had
led to the original suggestion that XTE~J1118$+$480 is a halo object,
probably as old as several Gyr~\cite{mir}. However, the recent
detection of high-Z elements in the optical spectrum of the source is
inconsistent with the hypothesis that it was formed from the direct
collapse of an old halo star~\cite{met}. On the other hand, the
metalicity, position, and spatial velocity of this binary system
requires a formation mechanism according to which it was produced in
the Galactic disk and was ejected from the disk by the asymmetric
explosion that formed the black hole~\cite{met,kick}. Even though this
formation history makes the black hole significantly younger than
estimated earlier, it also allows for an accurate bound to be placed
on its kinematic age.

Given the current position and 3D velocity of the source, it is
possible to integrate backwards its trajectory in the galactic
potential to infer the times at which it crossed the galactic
plane. Any of these crossings represents a plausible time at which the
black hole was formed. However, the last time at which the black hole
emerged from the galactic plane provides a lower limit on its
kinematic age. This approach has been used successfully in the past to
show that, in any of the recent crossing of the galactic plane, the
velocity imparted to the natal black-hole binary is consistent with
the kick velocities measured for other compact objects~\cite{kick}. In
this paper, I am only concerned with the time elapsed since the last
galactic crossing and not with the properties of the binary star
before or after the formation of the black hole.

\begin{table}[t]
\caption{\label{tab:table1}Observed kinematic properties of XTE~J1118$+$480}
\begin{ruledtabular}
\begin{tabular}{lcc}
 & Measurement & Uncertainty\\
\hline
Proper Motion, $\mu$ & 18.3~mas~yr$^{-1}$ & $\pm$1.6~mas~yr$^{-1}$ \\
Position Angle, $\theta$ & 246$^\circ$ & $\pm 6^\circ$\\
Radial Velocity, $V_{\rm r}$ & 10~km~s$^{-1}$ & $\pm$35~km~s$^{-1}$\\
Distance, $D$ & 1.72 kpc & $\pm$0.1 kpc\\
Mass, $M$ & 8.53~$M_\odot$ & $\pm$0.6~$M_\odot$
\end{tabular}
\end{ruledtabular}
\end{table}

There are four observable quantities that were recently measured
accurately for XTE~J1118$+$480 and allow for the reconstruction of its
evaporation history (see Table~\ref{tab:table1}). First, in order of
decreasing accuracy, are VLBA observations of the radio counterpart to
the source showed a measurable proper motion of $\mu=18.3\pm
1.6$~mas~yr$^{-1}$ along a position angle of $\theta=246\pm 6^\circ$.
Given a distance to the source, these numbers specify the magnitude
and direction of the source velocity on a plane perpendicular to our
line of sight.

Second, models of the orbital variation of the spectral lines observed
from the source led to a measurement of its radial
velocity~\cite{bh}. In the particular case of XTE~J1118$+$480, the
amplitude of orbital variations of the radial velocity ($\sim
500$~km~s$^{-1}$) is much larger than the reported systemic radial
velocity of the binary ($\sim$a few tens of km~s$^{-1}$), which is
also comparable to the measurement uncertainties ($\sim
20$~km~s$^{-1}$)~\cite{bh}.  This results in an uncertain measurement,
with the values reported ranging from $-15\pm 10$~km~s$^{-1}$ to
$+26\pm 17$~km~s$^{-1}$~\cite{bh}. Because of this large systematic
uncertainty, I will assume that the radial velocity is in the range
$(-25,+45)$~km~s$^{-1}$, which I will write approximately as $10\pm
35$~km~s$^{-1}$.

Finally, models of the optical/IR lightcurve of the source together
with a characterization of the spectrum of the companion star led to a
measurement of both the distance, $D$, to the source and of the masses
of the companion star and the black hole~\cite{unc}. Uncertainties
related to the contamination of the optical lightcurve by the
accretion disk~\cite{bh} and to the spectral characterization of the
companion star~\cite{unc} typically limit the accuracy of the
measurements. Here I use the rather optimistic values quoted in
reference~\cite{mass} rather than the more conservative estimates of
reference~\cite{bh}. Having made the opposite choice would have
affected only marginally the final result, the accuracy of which is
mostly determined by the uncertainty in the radial velocity
measurement.

A final ingredient in the calculation is the model for the galactic
potential in the vicinity of the sun. Because of the large number of
integrations needed in order to estimate the uncertainty of the
results (see below), I use here the simpler Paczynski potential for
the Galaxy~\cite{pac}. Changing the model parameters of the potential
within the range allowed by more recent studies of the galactic
kinematics in the solar vicinity~\cite{pot} affects the results by an
amount smaller than the uncertainty introduced by the systematics in
the radial velocity measurement. Given the model of the galactic
potential and the values for the observable properties of
XTE~J1118$+$480, I calculated the current 3D spatial velocity of the
source according to references~\cite{mir,transf} and performed the
calculation of the galactic crossing times according to
reference~\cite{pac}.

For the average values of the current kinematic parameters of
XTE~J1118$+$480 shown in Table~\ref{tab:table1}, the time elapsed
since the last galactic crossing is $\simeq$18~Myr. For a $8.5
M_\odot$ black hole, this corresponds to a limit on the asymptotic AdS
curvature of the extra dimensions (see eq.~\ref{eq:life}) of $L\le
0.06$~mm, which is tighter than the $L\le 0.2$~mm bound obtained by
current tabletop experiments of sub-mm gravity~\cite{TBE}. However,
because of the non-negligible uncertainties in the measurement of the
kinematic properties of XTE~J1118$+$480, and especially of its radial
velocity, I will now estimate the formal uncertainty on this limit.

I will assume that the measurements of the various kinematic
properties are independent of each other and assign to each one a
Gaussian probability distribution with a mean and standard deviation
equal to the values shown in Table~\ref{tab:table1}. The measurement
of the radial velocity is indeed independent of the measurement of the
distance and they are both independent of the measurements related to
the proper motion. On the other hand, the measurements of the proper
motion and of the corresponding position angle are correlated to some
level. However, because the uncertainties in these two parameters are
much smaller than the uncertainty in the radial velocity, it is safe
to neglect any correlated errors in them.

The assumption of Gaussian uncertainties is also easily justified for
the measurements of the proper motion and position angle, since they
are both dominated by statistical errors. However, the uncertainties
quoted in Table~\ref{tab:table1} for the radial velocity and distance
are dominated by systematic errors and rather represent a range of
allowed values. Making the assumption that these uncertainties are
also Gaussian leads to a more conservative calculation, since the
exponential wings of the Gaussian distribution function allow, in
principle, for a wider range of values for the measured quantities.

Given the assumptions discussed above, I can write the probability
that the current kinematic state of XTE~J1118$+$480 is described by a
proper motion between $(\mu,\mu+d\mu)$, a position angle between
$(\theta,\theta+d\theta)$, a radial velocity between $(V_r,V_r+dV_r)$,
and a distance between $(D,D+dD)$ as
\begin{eqnarray}
& &{\cal F}(\mu,\theta,V_r,D)d\mu d\theta dV_r dD =\nonumber\\
& & \qquad
F_\mu(\mu) d\mu F_\theta(\theta) d\theta F_{\rm V}(V_r)dV_r
 F_{\rm D}(D)dD\;,
\label{eq:prob}
\end{eqnarray}
where the probability distribution over each individual quantity is a
Gaussian. For each set of values for the kinematic properties, I can
now use the numerical method described above in order to calculate the
time $t_{\rm c}$ that has elapsed since the last crossing of the
galactic plane, which I will write formally as
\begin{equation}
t_{\rm c}=t_{\rm c}(\mu,\theta,V_r,D)\;.
\label{calc}
\end{equation} 
Given that the uncertainty in measuring the radial velocity is the
largest, I choose to make a change of variables in
equation~(\ref{eq:prob}) from $V_r$ to $t_{\rm c}$ and calculate the
probability distribution over $(t_{\rm c},\mu,\theta,D)$ as
\begin{eqnarray}
& &{\cal F}(t_{\rm c}\mu,\theta,D) dt_{\rm c}d\mu d\theta dD =
F_\mu(\mu) d\mu F_\theta(\theta) d\theta  F_{\rm D}(D)dD\nonumber\\
& & \qquad
\times F_{\rm V}[V_r(t_{\rm c},\mu,\theta,D)]
\left(\left.\frac{\partial t_{\rm c}}{\partial V_r}\right\vert_{\mu,\theta,D}
\right)^{-1}dt_{\rm c}\;,
\label{eq:change}
\end{eqnarray}
where $V_r(t_{\rm c},\mu,\theta,D)$ refers to the solution of
equation~(\ref{calc}) for the radial velocity, given a crossing time.
In evaluating equation~(\ref{eq:change}), I calculate numerically the
partial derivative using first-order finite differencing. Finally, I
marginalize over all the variables other than the crossing time to
obtain the probability distribution over crossing times.

\begin{figure}[t]
\epsfig{file=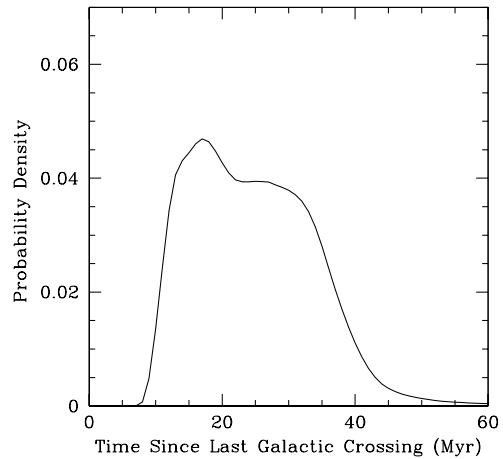, width=0.8\columnwidth}
\caption{\footnotesize
The probability distribution over time elapsed since the last
crossing through the galactic plane of the black hole XTE~J1118$+$480.
The median time is $\simeq 24$~Myr and the (95\%-probability) lower
limit is $\simeq 11$~Myr.}
\label{fig:prob}
\end{figure} 

\begin{figure}[t]
\epsfig{file=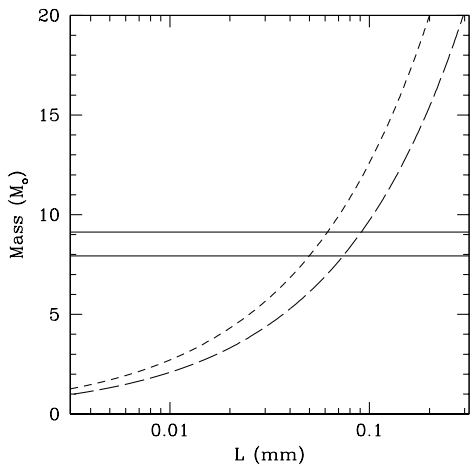, width=0.8\columnwidth}
\caption{\footnotesize
The curved lines show the constraints on the asymptotic AdS curvature
$L$ in the extra dimension placed by {\em (short dashed line)\/} the
most probable time since the last galactic crossing of XTE~J1118$+$480
and {\em (long dashed line)\/} the lower limit (at 95\% confidence) on
that time. The horizontal straight lines show the uncertainty in the
dynamical mass measurement of the black hole. The intersection of the
two gives the limit imposed by the kinematic age of the black hole of
$L>0.08$~mm.}
\label{fig:limits}
\end{figure}

The result is shown in Figure~\ref{fig:prob}. The lower limit
(95\%-probability) on the time since the last crossing of the galactic
plane is $\simeq $11~Myr, which is also the lower bound on the age of
the black hole. This translates on an upper limit on the asymptotic
AdS curvature of the extra dimensions of $L>0.08$~mm, as shown in
Figure~\ref{fig:limits}. Even though this limit is formally true only
for the particular RS2 braneworld model, it should also be similar, to
within factors of order unity, for any other model that has a finite
asymptotic curvature~\cite{emp}.

The limit on the asymptotic AdS curvature of the extra dimensions that
I derived above can be easily improved with a more accurate
measurement of the radial velocity of the binary system. Moreover,
measurement of the source position with VLBA during a subsequent
outburst will also reduce significantly the uncertainty in the proper
motion. The best case scenario would be the detection of a black hole
outside the galactic plane, traveling with a large velocity towards
the galactic plane. As an example, for a $4 M_\odot$ black hole, at
the same position as XTE~J1118$+$480, traveling with a velocity of
100~km~s$^{-1}$ towards the galactic plane, the minimum kinematic age
would be $\simeq 205$Myr and the upper bound on the asymptotic AdS
curvature of the extra dimensions would be $L<0.006$mm. This
represents the most stringent constraint that can be achieved with the
method discussed here.

{\em Note added:\/} After submission of this paper, Kapner et
al.~\cite{kap} reported an improved limit on the length scale in the
extra dimensions of $\le 50 \mu$m, which is comparable to the one 
obtained here.

%------------------------------------------------------

\bigskip
I thank Feryal \"Ozel and Keith Dienes for helpful comments on the
manuscript. I also thank Juan Maldacena and Nemanja Kaloper for useful
discussions, Scott Gaudi for help with current estimates of the
galactic potential, and the T-6 group of Los Alamos National Lab for
their hospitality.  This work was partially supported by NASA grant
NAG-513374.

\bibliographystyle{apsrev}

\end{document}